\documentclass[authoryear,12pt]{article}

\usepackage{latexsym}
\usepackage{graphicx}
\usepackage{graphics}
\usepackage{epsfig}
\usepackage{natbib}
\usepackage{times}
\usepackage{subfigure}
\usepackage{amssymb}
\usepackage{array}

\begin{document}

\title{Characterising the Anisotropic Mechanical Properties of Excised Human Skin}

\author{Aisling N\'{i} Annaidh$^{a,b}$, Karine Bruy\`{e}re$^c$, \\
Michel Destrade$^{a,d}$, Michael D. Gilchrist$^{a,e}$, M\'elanie Ott\'{e}nio\\[12pt]
$^a$School of Mechanical {\&} Materials Engineering, \\
University College Dublin, Belfield, Dublin 4, Ireland\\[12pt]
$^b$Institut Jean Le Rond D'Alembert, \\
UMR 7190, CNRS-UPMC, 4 place Jussieu, 75005 Paris, France\\[12pt]
$^c$ Ifsttar, LBMC, UMR\_T9406, F-69675, Bron, \\
Universit\'{e} Lyon 1, Villeurbanne, France\\[12pt]
 $^d$School of Mathematics, Statistics and Applied Mathematics,\\
  National University of Ireland Galway, Galway, Ireland\\[12pt]
$^e$School of Human Kinetics, \\
University of Ottawa, Ontario K1N 6N5, Canada}
\date{}

\maketitle

\newpage

\begin{abstract}
The mechanical properties of skin are important for a number of applications including surgery, dermatology, impact biomechanics and forensic science. In this study we have investigated the influence of location and orientation on the deformation characteristics of 56 samples of excised human skin. Uniaxial tensile tests were carried out at a strain rate of 0.012s$^{-1}$ on skin from the back. Digital Image Correlation was used for 2D strain measurement  and a histological examination of the dermis was also performed. The mean ultimate tensile strength (UTS) was 21.6$\pm$8.4MPa, the mean failure strain 54$\pm$17\%, the mean initial slope 1.18$\pm$0.88MPa, the mean ‘elastic modulus’ 83.3$\pm$34.9MPa and the mean strain energy was 3.6$\pm$1.6MJ/m$^3$. A multivariate analysis of variance has shown that these mechanical properties of skin are dependent upon the orientation of Langer lines (P$<$0.0001-P=0.046). The location of specimens on the back was also found to have a significant effect on the UTS (P =0.0002), the elastic modulus (P=0.001) and the strain energy (P=0.0052). The histological investigation concluded that there is a definite correlation between the orientation of Langer Lines and the preferred orientation of collagen fibres in the dermis (P$<$0.001). The data obtained in this study will provide essential information for those wishing to model the skin using a structural constitutive model.
\end{abstract}



\section{Introduction}

Skin is a complex multi-layered material which can broadly be divided into three main layers, the epidermis, the dermis, and the hypodermis. The epidermis consists of cells and cellular debris, the dermis consists of mostly networks of the fibrous proteins collagen, reticulin and elastin \citep{Wilkes73}, and the hypodermis is primarily made up of connective tissue and fat lobules. Collagen fibres account for 75\% of the dry weight of dermal tissue \citep{Wilkes73} and it is these fibres which are responsible for the strength of the skin.

Skin is a highly non-linear, anisotropic, viscoelastic and nearly incompressible material. A typical stress-stretch graph for skin exhibits non-linear behaviour and its response is usually described as three-phase \citep{Brown73}. In the initial loading phase, skin is very compliant and large deformation occurs at a relatively low applied load. At this stage the fibres are largely unaligned. In the second phase the stiffness of the skin gradually increases as the fibres align themselves in the direction of applied load. The third phase is an almost linear phase whereby the stiffness increases rapidly as the collagen fibres are mostly aligned and the overall mechanical response becomes dependent upon the mechanical properties of the collagen fibres, which are three grades of magnitude stiffer than elastin.

That skin is anisotropic was recognized as far back as the 19th century by Karl \citet{Langer78a}, who mapped the natural lines of tension which occur within the skin. These lines are identified by puncturing the skin with a circular device. The wounds then assume an elliptical shape and by joining the major axes of the ellipses a system of lines can be drawn as shown in Fig.~\ref{samples}(a). These lines are known as the Langer lines. While Langer lines are the best known skin tension lines, many variations on the original lines proposed by Langer have been made. These include the Cox lines \citep{Cox41}, Kraissl lines \citep{Kraissl51} and Relaxed Skin Tension Lines \citep{Borges84}. 

Early tensile tests suggested that the deformation characteristics of skin are dependent upon specimen orientation with respect to the Langer lines \citep{Ridge66a}. More recent work conducted using Optical Coherence Elastography in-vivo, indicates a large difference between the Young moduli of skin along directions parallel and orthogonal to the Langer lines \citep{Liang}. In this study, we provide new experimental data of in-vitro human skin focusing in particular on the anisotropic properties of the skin. We also quantify the degree of anisotropy of the mechanical properties with respect to their Langer line orientation. Finally, we perform a quantitative histological analysis on the dermis of the test samples to examine a possible correlation between the orientation of Langer lines and the preferred direction of collagen fibres.

While many recent publications on skin report in-vivo experiments \citep{Delalleau08a, Flynn11, Moerman09}, here we have limited our experiments to in-vitro uniaxial tensile tests.  As discussed by Holzapfel, there are a number of issues involved with tensile tests \citep{Holzapfel06b}: The structural integrity of the samples may be disturbed at the lateral edges and due to the soft nature of the test material it is difficult to insert samples without subjecting them to an axial load. Despite these shortcomings, it was decided to perform in-vitro tensile tests over in-vivo testing for a number of reasons.  First, in-vitro tests provide simple stress-strain relationships that can be modeled and quantified easily, since their boundary conditions are well defined. Second, for the application of this work in a further study into the dynamic rupture of skin, we are also interested in failure and rupture of skin, and in-vitro testing makes it possible to test until failure.

Most in-vitro investigations use human skin substitutes such as pigskin, silicone or polyurethane \citep{McCarthy07, Gilchrist08, Jor11, McCarthy10} and in-vitro tests on human skin are particularly rare. Those classical publications which do use human skin often rely on outdated equipment and their results can be difficult to interpret \citep{Cox41, Jansen58b}. However, the development of accurate constitutive models depends heavily upon their material definitions. This study aims to provide new material data for human skin which can be applied to constitutive models in a number of technical areas such as razor blade manufacture, cosmetics, surgical simulation, forensic pathology and impact biomechanics. Uniaxial tensile tests alone are not sufficient to determine multi-dimensional material models for soft tissues and using a simple non-linear regression analysis to determine constitutive parameters may result in non-unique solutions, ill-conditioned equations and slow convergence rates. Ideally, testing of skin would involve planar biaxial tests with an in-plane shear, and separate through-thickness shear tests. While advanced testing protocols like these are not yet available, simple tension tests remain important because they serve to evaluate the level of anisotropy and to provide data which can later be used as validation for models constructed using more complicated testing methods \citep{Holzapfel09}. Moreover, model parameters can be provided directly from a histological study of the collagen fibre alignment in the dermis which allows for reasonable determination of material responses \citep{Holzapfel06b}.


\section{Materials {\&} Method}


\subsection{Specimen Preparation}
\label{SpecimenPreparation}

Tensile tests and preparation were carried out in Ifsttar (Institut fran{\c c}ais des sciences et technologies des transports, de l'am\'{e}nagement et des r\'{e}seaux), France. French law allows human corpses that have been donated to science to be used for research purposes. The ethics committee within Ifsttar approved the use of human biological material. Skin was excised from seven subjects: three male, four female. The average age of the subjects was 89$\pm$6 years and none had any related skin diseases. Table \ref{details} provides details on the gender and age of each subject at the time of death along with the number of days that elapsed prior to testing.

\newcolumntype{x}[1]{%
>{\centering\hspace{0pt}}p{#1}}%
\begin {table}[!ht]
\centering
\caption{Information on each test subject. \textasteriskcentered Number of days elapsed after death prior to testing \label{details}}
\begin{tabular}{|x{2.5cm}|x{2.5cm}|x{2.5cm}|x{2.5cm}|}
\hline
Gender		& Age (years)	&Time Elapsed\textasteriskcentered	& Label \tabularnewline
\hline
Male		& 97		& 9		& A \tabularnewline
Male		& 91		& 3		& B \tabularnewline
Female		& 97		& 2		& C \tabularnewline
Female		& 81		& 3		& D \tabularnewline
Male		& 89		& 3		& E \tabularnewline		
Female		& 85		& 4		& F \tabularnewline
Female		& 86		& 4		& G \tabularnewline
\hline
\end{tabular}
\end{table}

\begin{figure}[!ht]
\centering
\subfigure[]{\includegraphics[trim = 25mm 2mm 0.5mm 2mm, clip, scale=0.125]{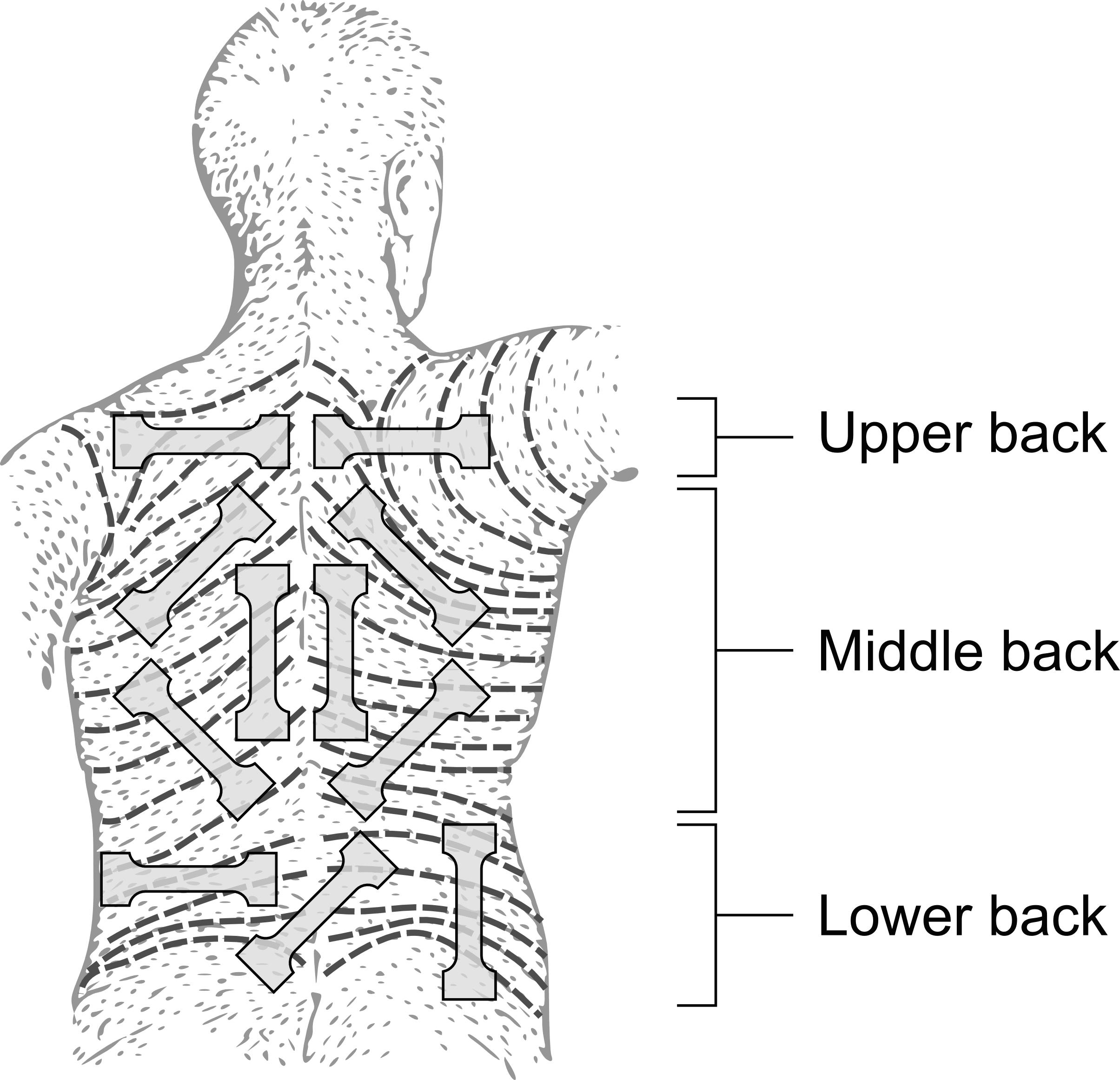}}
\subfigure[]{\includegraphics[trim = 0mm 60mm 30mm 5mm, clip,scale=0.5,angle=90]{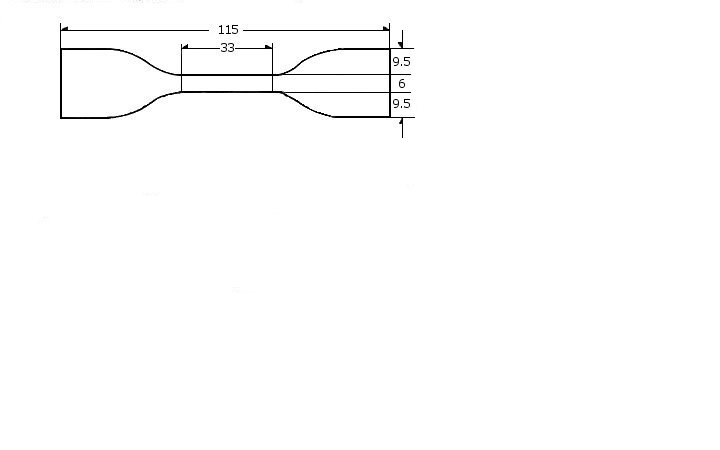} }
\caption{(a) Orientation of samples from the back. Specimens are superimposed onto a map of the Langer lines to indicate their orientation with respect to the lines \citep{Langer78a}. (b) Dimensions of custom made die (mm) \label{samples}}
\end{figure}

Only the skin from the back of the subject was available for use. This was excised from the body with a scalpel. Each sample of skin was cut into a dogbone shape specimen, according to the ASTM D412 Standard tensile test method for vulcanised rubber, using a custom made die (see Fig.~\ref{samples}). Each specimen then had the epidermis and any underlying adipose tissue carefully removed with a scalpel. The thickness of the skin after removal of adipose tissue was measured using a Vernier Calipers and the mean thickness was 2.56$\pm$0.39mm. It was observed that skin removed from the lumbar area of the back contained more adipose tissue than skin form other areas of the back. The dimensions of the test specimens were measured before and after excision. Specimens were obtained in various orientations, shown in Fig.~\ref{samples}, to correlate the specimens with the direction of Langer lines. Each sample was grouped into one of three categories: parallel, perpendicular, or at 45$^\circ$ to the Langer lines. The skin was stored in moistened paper and refrigerated at 4$^\circ$C until it was ready to be tested. A total of 56 samples were tested successfully. Although it is widely accepted that preconditioning is necessary at small and intermediate strains, at higher strains the stress-strain response is not significantly affected by preconditioning \citep{Eshel01}. The samples were therefore not pre-conditioned prior to testing.

\subsection{Tensile Tests}
\label{Tests}

Tensile tests were performed using a Universal Tensile Test machine. Samples were clamped using specially designed anti-slip clamps, to counteract the tendency of samples to slip in ordinary grips. The velocity of the cross-head was 50mm/min. The tensile load was measured with a 1kN piezoelectric load cell. Each tensile test was videoed with two Dalsa Falcon 1.4M100 digital video cameras at 20 frames/second. This was to record any abnormal behaviour during the experiments and also for the subsequent use of Digital Image Correlation. The gauge length and width were both measured optically. For these tests, the strain rate was 0.012s$^{-1}$, nominally the same strain rate as in tests carried out by \citet{Jansen58a}. 

The main focus of this study is an investigation into the hyperelastic properties of skin and how these vary according to orientation and location. We thus considered the tensile tests to be taking place at quasi-static speed, and ignored viscoelastic effects for the purpose of constitutive hyperelastic modelling (for examples of studies where this route is chosen, see \citet{Jor11}, \citet{Pailler08}, \citet{Evans09b} or \citet{Cavicchi09}). Further creep and stress relaxation tests, requiring additional skin samples, would need to be carried out to characterize the viscoelasticty of skin.


\subsection{Digital Image Correlation}

The stretch ratio was calculated via the displacement cell attached to the cross-head of the tensile machine. As validation, the stretch ratio was also calculated via Digital Image Correlation (DIC). DIC is a full-field optical strain measurement technique which uses image registration to measure the 2D or 3D deformation of a material. The technique works by tracking unique features on the surface of the material as it deforms. The correlations were based on images taken with the two Dalsa Falcon video cameras and processed using Vic2D ® Software (Version 2009-Correlated Solution, Inc.). Black spray paint was applied to the surface of the skin to create the desired random speckled pattern necessary for the correlation.


\subsection{Histology}

Three biopsies were procured from the test specimens prior to testing. These biopsies were placed in a Formaldehyde solution for fixation for a period exceeding 48 hours. Biopsies were then embedded in paraffin prior to slicing. Each biopsy was sectioned at 5$\mu$μm intervals, creating a number of thin slices which were transferred onto glass slides. Biopsies were sliced in three orthogonal planes as shown in Fig.~\ref{biopsies} below.
The samples were then stained using a Van Gieson stain, which differentiates collagen fibres by making them appear pink/red. Images were taken of each slide using a Nikon E80i Transmission microscope and CCD camera. The orientation of collagen fibres were then calculated in a fully-automated customized MATLAB routine using the Image Processing Toolbox. The algorithm is described further in \citep{NiAnnaidh11b}.

\begin{figure}[!ht]
\centering
 \includegraphics[trim = 0mm 80mm 10mm 0mm, clip,scale=0.75]{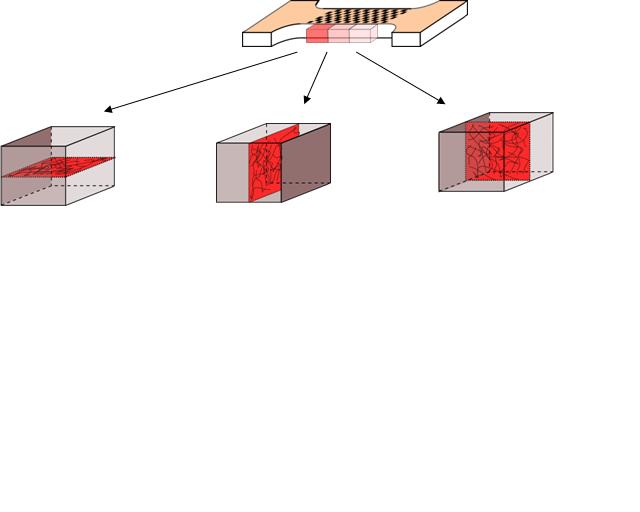}
\caption{Biopsies of skin samples for purpose of histological staining. Note that the biopsies have been sliced in three orthogonal planes.}
\label{biopsies}
\end{figure}


\subsection{Statistical Analysis}

A statistical analysis of the experimental results was carried out using the General Linear Model procedure implemented in SAS 9.1 (SAS Institute Inc., USA). A multivariate analysis of variance was utilised, followed by the Tukey-Kramer post-hoc test, to determine the influence of orientation and location of the various test samples. Significance levels for these tests were set to P$<$0.05. When performing this test a normal distribution was assumed. To ensure our data set followed a normal distribution, a Lilliefors test was performed using the \emph{lillietest} function in MATLAB R2007b.


\section{Results}

\begin{figure}[!ht]
\centering
\includegraphics[trim = 40mm 10mm 40mm 15mm, clip, scale=0.28]{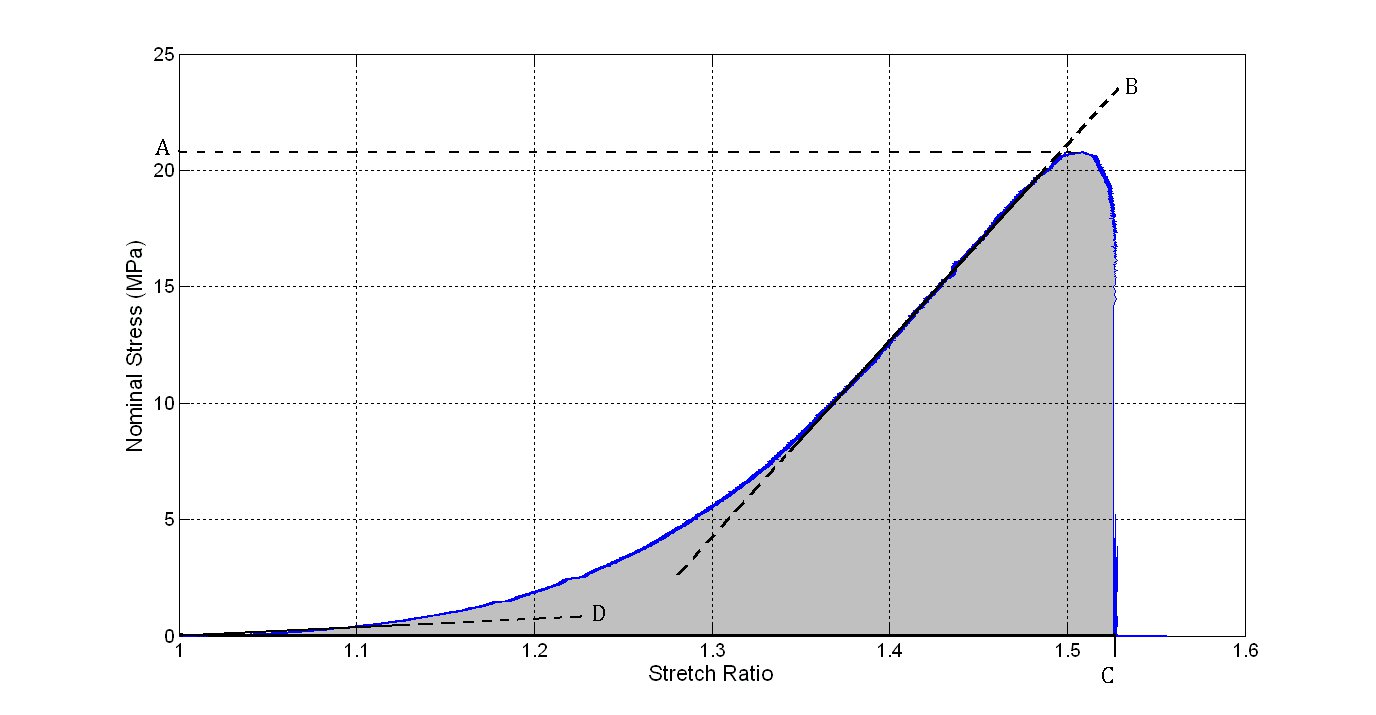}
\caption{Typical stress-stretch graph for the experiments. The ultimate tensile strength is the maximum stress until failure of the specimen and is indicated by A. The “elastic modulus” is defined as the slope of the linear portion of the curve shown by B. The failure stretch is the maximum stretch obtained before failure and is shown by C. The initial slope is the slope of the curve at infinitesimal strains and is shown by D. The strain energy is the energy per unit volume consumed by the material during the experiment and is represented by the area under the curve.\label{typical}}
\end{figure}

For each tensile test performed a force-displacement curve was obtained. The nominal stress was then calculated by dividing the force by the undeformed cross sectional area of the specimen. The stretch ratio was calculated by dividing the current length of the specimen by the initial length. In this way nominal stress Vs stretch ratio graphs were plotted for each specimen. A number of characteristics from these curves were identified as descriptive parameters. They are illustrated in Fig.~\ref{typical}. 

\subsection{Intra-Subject and Inter-Subject Variation}

\begin{figure}[!ht]
\centering
\subfigure[]{\includegraphics[trim = 55mm 5mm 25mm 70mm, clip, scale=0.45]{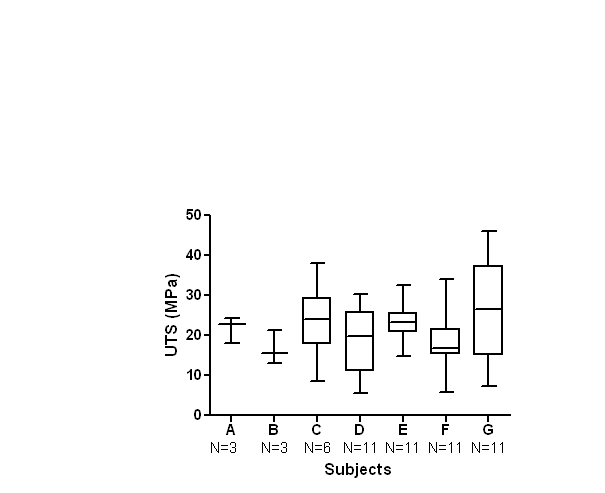}}
\subfigure[]{\includegraphics[trim = 25mm 15mm 30mm 15mm, clip, scale=0.27]{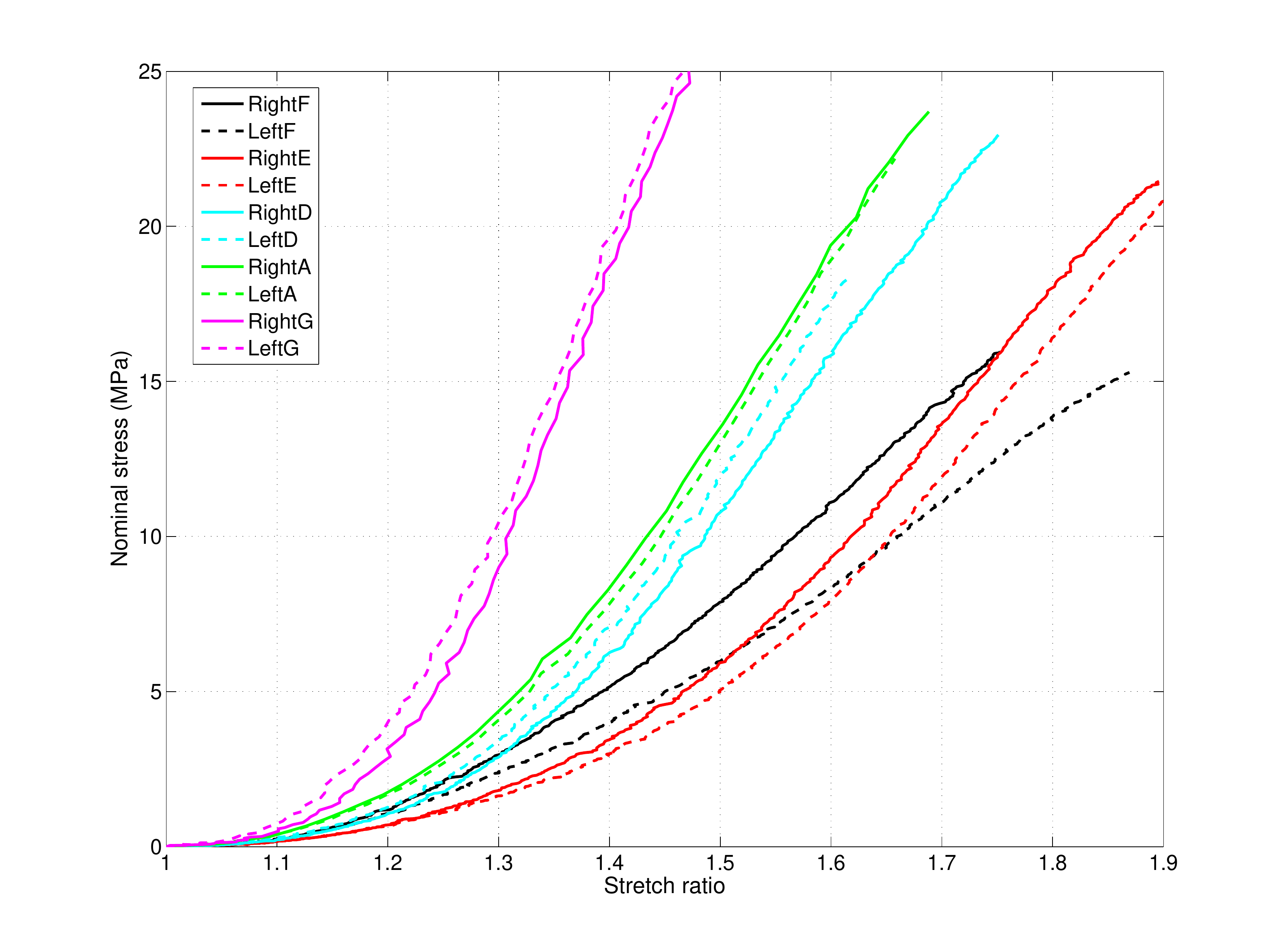} }
\caption{(a) Box and whisker plot of the intra-subject and inter-subject variation in UTS (N=56) for subjects A-G. The whiskers indicate the max and min. The boxes represent the lower and upper quartile and the line is the median. (b) Stress-Stretch response of adjacent samples from the upper back (see Fig.~\ref{samples}) with the same orientation (N=10). Colours indicate different subjects while solid and dashed lines indicate samples to the right and left of the spinal line respectively. This illustrates the inter-subject and intra-subject variation between adjacent samples with the same orientation. \label{SubjectVariation}}
\end{figure}

A difficulty often encountered with the testing of biological tissue is the large variation in experimental results across samples. Fig.~\ref{SubjectVariation}(a) indicates the intra-subject and inter-subject variation in the UTS between samples. There is a large intra-subject variation for each subject which is primarily due to anisotropy. Fig.~\ref{SubjectVariation}(b) shows the stress-stretch graphs of ten different samples taken from five subjects. Note that the left and right samples were adjacent to one another but also in the same orientation with respect to the Langer lines. The similarity of these responses between the left and right samples from the same subjects indicates good repeatability of our experiments.

\subsection{Derformation of Specimens After Excision}

It is known that both shrinkage and expansion of specimens can occur upon excision from the body \citep{Cox41, Jansen58a, Ridge66a, Lanir74}. This is due to the release of residual stresses within the skin. What remains unclear is the level of residual stress present within the skin and how this may vary with the orientation of specimens. It has been suggested that the shrinkage of excised specimens is greatest in the direction of Langer lines \citep{Ridge66a} however our results did not corroborate this. Our results showed that 73\% of all samples expanded upon excision, while only 27\% of all samples shrunk. Of the samples that expanded there was a mean expansion of 5.1$\pm$3.4\%. Amongst the samples that shrunk, a mean shrinkage of 3$\pm$2.9\% was recorded. This is far below the reported maximum figure of 20\% shrinkage for human skin \citep{Jansen58a} and 40\% shrinkage provided for pig skin \citep{Jor11}.


\subsection{Histology}

The collagen fibres were assumed to form an interweaving lattice structure, as first postulated by \citet{Ridge66a}. Here, the preferred orientation is defined as the mean of the orientations of two families of interweaving fibres. It was assumed that the collagen fibres are distributed according to a transversely isotropic and $\pi $-periodic Von Mises distribution. The Von Mises distribution is commonly assumed for directional data. The standard $\pi $-periodic Von Mises Distribution is normalized and the resulting distribution is given as follows.

\begin{equation}
\rho(\Theta)=4\sqrt{\frac{b}{2\pi}}\frac{\exp[b(\cos(2\Theta)+1]}{erfi\sqrt{2b}}
\label{Von mises}
\end{equation}
where $\Theta $ is the mean orientation of fibres  and $b$ is the concentration parameter associated with the Von Mises distribution. The parameter $b$ was evaluated by the approximation given by \citet{Fisher93} using the CircStat package available on the MATLAB file exchange \citep{Berens09}.

Table~\ref{HistoOrientations} shows the calculated concentration parameter $b$, the mean orientation of collagen fibres and the orientation with respect to Langer lines. A Pearson correlation test was carried out to test for a correlation between the measured preferred orientation obtained through histology and the perceived orientation of Langer lines. The correlation was deemed to be significant (P$<$0.001) with a Pearson R value of 0.9487 and an R$^2$ value of 0.9000.

\newcolumntype{x}[1]{%
>{\centering\hspace{0pt}}p{#1}}%
\begin {table}[!ht]
\centering
\caption{Mean orientation of collagen fibres and their Langer Line orientations. (Note that the data given is axial data i.e. it represents undirected lines and does not distinguish between $\theta $ and $\pi +\theta$ \citep{Jones06} e.g. the orientation of 0$^\circ$ and 180$^\circ$ are equivalent).\label{HistoOrientations}}
\begin{tabular}{|x{2.5cm}|x{2.5cm}|x{2.5cm}|}
\hline
Orientation with respect to Langer Lines	& Preferred Orientation, $\Theta$  &Level of dispersion, $b$\tabularnewline
\hline
0$^\circ$		& 40$^\circ$		&0.5222 \tabularnewline
0$^\circ$		& 163$^\circ$		&1.1605\tabularnewline
0$^\circ$		& 16$^\circ$		 &1.1471\tabularnewline
0$^\circ$		& 168$^\circ$		&0.6004 \tabularnewline
45$^\circ$		& 40$^\circ$		&0.8388 \tabularnewline
45$^\circ$		& 61$^\circ$		&0.7926 \tabularnewline
45$^\circ$		& 46$^\circ$		 &0.7453\tabularnewline
45$^\circ$		& 53$^\circ$		 &0.4209\tabularnewline
45$^\circ$		& 57$^\circ$		 &0.8326\tabularnewline
45$^\circ$		& 83$^\circ$		 &0.9330\tabularnewline
90$^\circ$		& 73$^\circ$		 &0.9007\tabularnewline
90$^\circ$		& 120$^\circ$		 &0.6197\tabularnewline
\hline
\end{tabular}
\end{table}


\subsection{Influence of Orientation}

\begin{figure}
\centering
 \includegraphics[trim = 15mm 40mm 15mm 45mm, clip, scale=0.5]{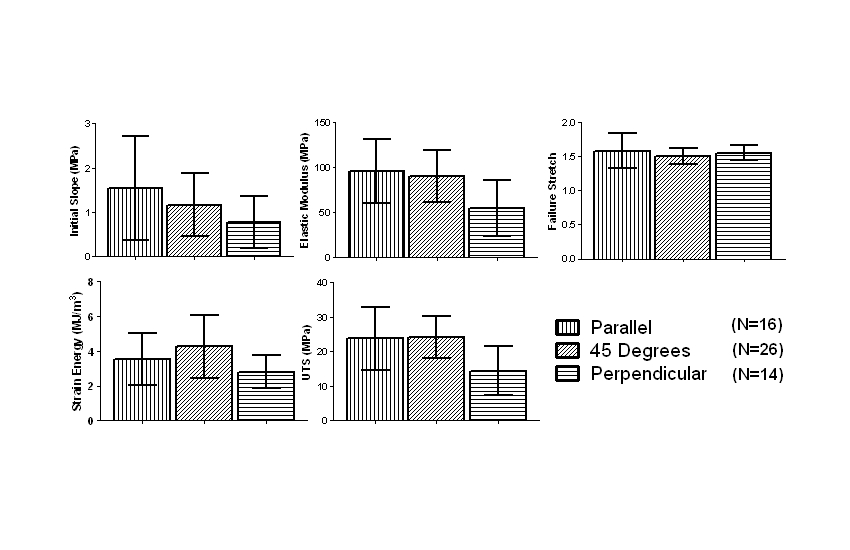}
\caption{Influence of orientation on the inital slope, elastic modulus, failure stretch, strain energy and UTS. Values given include mean and standard deviations \label{orientation}}
\end{figure}

Specimens were analysed by considering their orientation with respect to the Langer lines (parallel, perpendicular or at 45$^\circ$ to the Langer lines, see Fig.~\ref{samples}). A multiway analysis of variance found the orientation of Langer lines to have a significant effect on the UTS (P$<$0.0001), the strain energy (P=0.0101), the elastic modulus (P=0.0002), the initial slope (P=0.0375), and the failure stretch (P=0.046). The interaction between orientation and location was also tested i.e., whether the effect of orientation was dependent upon location. This interaction between orientation and location was found to be significant only for the failure stretch (P=0.0118). The presence of this interaction is probably due to small differences in the perceived orientation of Langer lines over the body. Fig.~\ref{orientation} and Table~\ref{summary} display the variation in the mechanical properties between samples of different orientations. The variations are substantial, except for the failure stretch, where the effect of the orientation is complicated by the existence of the interaction between orientation and location.

\newcolumntype{x}[1]{%
>{\centering\hspace{0pt}}p{#1}}%
\begin {table}[!ht]
\centering
\caption{ Mean $\pm$ standard deviation for each orientation/location group.\label{summary}}
\begin{tabular}{|x{2cm}|x{1.5cm}|x{0.8cm}|x{1.5cm}|x{1.5cm}|x{1.5cm}|x{1.5cm}|x{1.5cm}|}
\hline
Langer Line Orientation	&Location	&N	&UTS (MPa)	  	&Strain Energy (MJ/m$^3$)	&Failure Stretch	&Elastic Modulus (MPa)	&Initial Slope (MPa)		\tabularnewline
\hline
Parallel				& Middle	& 9	& 28.64 $\pm$\footnotesize{9.03}  	&4.28 $\pm$\footnotesize{1.49}	&1.46 $\pm$\footnotesize{0.07}	&112.47 $\pm$\footnotesize{36.49}	&1.21 $\pm$\footnotesize{0.97}\tabularnewline
Parallel				& Bottom	& 7	& 17.60 $\pm$\footnotesize{4.77}  	&2.54 $\pm$\footnotesize{0.76}	&1.74 $\pm$\footnotesize{0.32}	&73.81 $\pm$\footnotesize{19.41}	&1.95 $\pm$\footnotesize{1.34}\tabularnewline
45$^\circ$			& Top		& 12	& 22.7 $\pm$\footnotesize{3.61}  	 &3.80 $\pm$\footnotesize{0.92}	&1.52 $\pm$\footnotesize{0.10}	&82.62 $\pm$\footnotesize{17.36}	&0.99 $\pm$\footnotesize{0.51}\tabularnewline
45$^\circ$			& Middle	& 9	& 28.85 $\pm$\footnotesize{7.87}  	&5.38 $\pm$\footnotesize{2.59}	&1.52 $\pm$\footnotesize{0.15}	&103.49 $\pm$\footnotesize{41.20}	&1.33 $\pm$\footnotesize{0.96}\tabularnewline
45$^\circ$			& Bottom	& 5	& 20.23 $\pm$\footnotesize{4.18} 	&3.31 $\pm$\footnotesize{0.67}	&1.43 $\pm$\footnotesize{0.04}	&82.81 $\pm$\footnotesize{18.43}	&1.30 $\pm$\footnotesize{0.60}\tabularnewline	
Perpendicular			& Middle	& 9	& 16.53 $\pm$\footnotesize{5.71} 	&2.98 $\pm$\footnotesize{0.89}	&1.52 $\pm$\footnotesize{0.08}	&63.75 $\pm$\footnotesize{24.59}	&0.91 $\pm$\footnotesize{0.68}\tabularnewline
Perpendicular			& Bottom	& 5	& 10.56 $\pm$\footnotesize{8.41}	&2.44 $\pm$\footnotesize{1.04}	&1.61 $\pm$\footnotesize{0.14}	&37.66 $\pm$\footnotesize{36.41}	&0.54 $\pm$\footnotesize{0.33}\tabularnewline
\hline
\end{tabular}
\end{table}

\subsection{Influence of Location}

\begin{figure}
\centering
 \includegraphics[trim = 15mm 5mm 15mm 10mm, clip, scale=0.4]{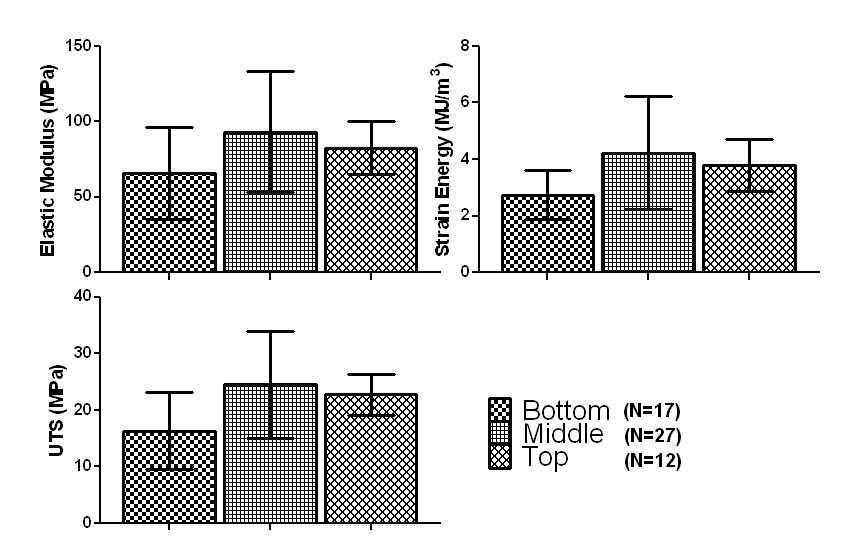}
\caption{Influence of location on back on the elastic modulus, strain energy and UTS. Values given include mean and standard deviations. \label{location}}
\end{figure}

When results were grouped into three locations (Upper Back, Middle Back and Lower Back, see Fig.~\ref{samples}), a multiway analysis of variance also found the location of specimens to have a significant effect on the UTS (P=0.0002), the strain energy (P=0.0052) and the Elastic Modulus (P=0.001), but neither on the failure stretch nor the initial slope. Fig.~\ref{location} displays variations in the mechanical properties between different locations on the back for the three properties which showed a significant effect i.e., the elastic modulus, UTS and strain energy. Note that there was no significant effect noted for either the initial slope or the failure stretch.

\subsection{Digital Image Correlation}

\begin{figure}[!ht]
\centering
\subfigure[]{\includegraphics[trim = 1mm 10mm 1mm 10mm, clip, scale=0.575]{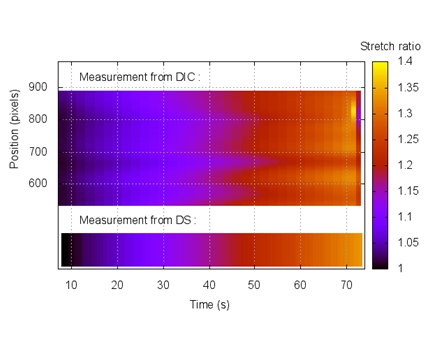}}
\subfigure[]{\includegraphics[trim = 50mm 1mm 50mm 10mm, clip, scale=0.6]{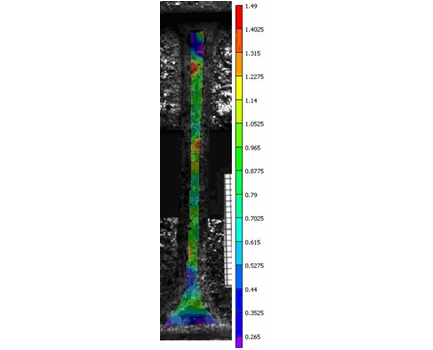} }
\caption{(a) Comparison of stretch measurement from DIC and DS over the length of the specimen throughout the duration of a given tensile test. (b) Distribution of local Lagrangian strains just before rupture. Note the area of localised magnification in red. This is the location where rupture subsequently occurred.\label{DIC}}
\end{figure}

An important feature of DIC is its ability to calculate local strains throughout the plane of the test specimens. Fig.~\ref{DIC}(a) shows a comparison between the local stretch values, obtained through DIC, and the overall stretch values, obtained from the displacement sensor (DS), throughout the duration of a test.  It can be seen that there is good agreement for most of the duration of the test.  However, upon closer inspection it was found that a local magnification of the stretch ratio occurred just before rupture, close to the rupture site (for magnitudes, see Table~\ref{LocalStretch}).  Fig.~\ref{DIC}(b) shows the distribution of strains throughout the sample for a frame just before rupture.  For some samples this magnification was very localised while for others it was distributed over a larger area.  A similar result was reported by  \citet {Jacquemound07b} who found that at the rupture site, the local strain values can be up to twice as large as what is measured using overall specimen strains calculated over the entire length of a specimen.

\newcolumntype{x}[1]{%
>{\centering\hspace{0pt}}p{#1}}%
\begin {table}[!ht]
\centering
\caption{Local and overall maximum stretch ratios for five different samples. Local magnification of the stretch ratio occurs just before rupture. \label{LocalStretch}}
\begin{tabular}{|x{2.5cm}|x{2.5cm}|x{3.2cm}|}
\hline
Sample		& Local Max Stretch Ratio	&Overall Specimen Stretch Ratio \tabularnewline
\hline
1		&1.47				&1.38 \tabularnewline
2		&1.45				&1.40\tabularnewline
3		&1.65				&1.45\tabularnewline
4		&1.70				&1.48\tabularnewline
5		&1.70				&1.50\tabularnewline
\hline
\end{tabular}
\end{table}


\section{Discussion}

When comparing our own data to that in the literature it can be seen that there is much variation between authors. This is to be expected for biological tissues. First, because of the biological variability between subjects; second, because of the anisotropic nature of skin itself; and third,  because of the sensitivity of biological tissues to test conditions. In the absence of extensive published literature on the properties of in vitro human skin from the back, Table~\ref{comparison} compares results obtained from the literature from a range of locations and test methods to our own results. The mean ultimate tensile strength was 21.6$\pm$8.4MPa, the mean failure strain 54$\pm$17\%, the mean initial slope 1.18$\pm$0.88MPa, the mean elastic modulus 83.3$\pm$34.9MPa and the mean strain energy was 3.6$\pm$1.6MJ/m$^3$. It can be seen that our experimental results fall within the ranges found in the literature for in-vitro tests. However there is a large variation with the various types of in-vivo tests. The reasons for this difference has been widely discussed and the primary causes are due to the properties of skin changing once removed from the body; the different testing methods employed; and an inability to establish boundary conditions and to isolate the skin's behaviour from its surrounding tissue in vivo. Furthermore, the 'Modulus' defined in-vivo is in the low-strain range, whereas the Modulus calculated here is in the high-strain range. The in-vivo modulus then, should be compared with the Initial slope, rather than the elastic modulus.

\begin {table}[!ht]
\centering
\caption{Comparison of results obtained to results in the literature. All experiments were carried out in-vitro on excised human skin. Results are displayed as means ($\pm$standard deviation) and ranges.\label{comparison}}
\begin{tabular}{|x{2.2cm}|x{1.8cm}|x{1cm}|x{1.5cm}|x{1.5cm}|x{1.5cm}|x{1cm}|x{1cm}|}
\hline
Author 		&Test type 				&UTS (MPa)		&Failure Strain \%	&Elastic Modulus MPa		&Initial Slope MPa 		&Site\textasteriskcentered 		&Age 				\tabularnewline
\hline
\citep{Jansen58b}		&in vitro tension			&1-24			&17-207		&2.9-54.0			&0.69-3.7			&Ab					&0-99				\tabularnewline\hline		
\citep{Dunn83}		&in vitro tension			&2-15			&			&18.8				&0.1				&AB\&T				&47-86				\tabularnewline\hline	
\citep{Vogel87}		&in vitro tension			&5-32			&30-115		&15-150			&				&V					&0-90				\tabularnewline\hline	
\citep{Jacquemound07b}	&in vitro tension			&5.7-12.6		&27-59			&19.5-87.1			&				&F \& A				&62-98				\tabularnewline\hline	
\citep{Agache80}		&in vivo torsion			&			&			&0.42-0.85			&				&Back					&3-89				\tabularnewline\hline	
\citep{Diridollou98}	&in vivo suction			&			&			&0.12-0.25			&				&A\&F					&20-30				\tabularnewline\hline	
\citep{Khatyr04}		&in vivo tension			&			&			&0.13-0.66			&				&Tibia					&22-68				\tabularnewline\hline	
\citep{Pailler08}	&in vivo indentation			&			&			&0.0045-0.008		&				&A					&30				\tabularnewline\hline	
\citep{Zahouani09}	&in vivo indentation \& static friction			&			&			&0.0062-0.0021		&				&A					&55-70				\tabularnewline\hline	
\hline
Our results		&in vitro tensile			&21.6 $\pm$8.4	&54 $\pm$17		&83.3 $\pm$34.9		&1.18 $\pm$0.88		&Back					&81-97				\tabularnewline	
\hline
\end{tabular}
\textasteriskcentered \footnotesize{Ab = abdomen, T = thorax, V = various, F = forehead, A = arm}
\end{table}

It has been well documented over the years that the deformation characteristics of skin are dependent upon specimen orientation, but few authors describe the level of anisotropy involved in much detail. In this study, we have provided statistical information on the effect of Langer line orientation on the UTS, strain energy, failure strain, initial slope, and elastic modulus. We have been unable to establish a link between the level of shrinkage of skin upon excision and the orientation of the Langer lines. Instead we found that the majority of samples actually expanded upon excision. The reason for this discrepancy may be two-fold; firstly, the skin has been excised from the back, not the abdomen, which may be more extensible to accommodate the shrinkage and expansion of the stomach. Secondly, the skin samples were excised from elderly subjects which naturally exhibit less extensibility. This has been previously noted by \citet{Jansen58a, Jansen58b} who stated that skin excised from elderly subjects was more likely to expand rather than retract, when compared to younger subjects. Finally, the authors noted that while shrinkage of the skin may have occurred directly after excision from the body, after subcutaneous fat was removed, the samples often expanded. This poses the question what effect the subcutaneous fat has on the level of shrinkage/ expansion in the skin, usually thought to be solely due to the release of residual stresses in the skin.

Previous studies show that the properties of skin are dependent upon specimen location \citep{Haut89, Sugihara91}. Our results also indicate that the UTS, strain energy and elastic modulus vary depending on location. However, as only skin from the back was available for testing, the location of specimens could only be categorised into upper back, middle back and lower back regions. It was nonetheless advantageous to gain information on skin from the back because it is seldom as readily available as that from the abdominal area.

All subjects were between the ages of 81 and 97. This meant that no comparisons based on the age of subjects could be performed. Numerous other studies have investigated the influence of age on the structure and mechanical behaviour of skin \citep{Vogel87, Haut89, Sugihara91}. It has been established that progressive straightening of elastin fibres occurs over time \citep{Lavker87}. This may cause the shortened phase I region of the stress-strain curve leading to an overall reduction in extensibility. The slope of the phase III portion of the curve however remains relatively constant \citep{Wilkes73}. Nonetheless, it must be highlighted that the results of this study may not reflect the behaviour of younger skin and consideration must be taken to allow for age-related effects when interpreting these results.

A number of days had elapsed between the time of death and testing (see Table~\ref{details}) but it has been previously demonstrated that skin samples can be refrigerated for up to 48 hours under the right conditions with no discernible effects on the mechanical properties \citep{Daly82}  and for up to 21 days for thoracic aortas under certain conditions \citep{Adham96}.

DIC is a common strain measurement technique in biomechanics, however until now it has been used mostly for in-vivo experiments or at low strains \citep{Moerman09}. The results yielded from the present use of DIC may be useful in the future as validation for a finite element approach. Also the reported local maximum strain values are of interest to those wishing to model failure of biological tissues.

In this paper, our data has been examined with respect to the perceived orientation of Langer lines. However, the Langer lines are known to vary with location, age and gender and no universal pattern of maximum tensions exist \citep{Brown73}. The orientation of the Langer lines cannot be identified with certainty unless the skin of a whole cadaver is punctured, an option which was not feasible for this study. Given the importance of the Langer lines on the mechanical properties of skin there is a need for patient-specific maps to be established in-vivo and in real time. These could be produced by relying on sophisticated imaging techniques such as those produced by ultrasonic surface wave propagation or optical coherence tomography and may eventually have applications in cosmetic surgery.

\citet{Cox41} and \citet{Stark77}  both illustrated that the orientations of Langer lines remain even after the skin is removed from the body and the skin tension released, and concluded that the lines have an anatomical basis. However it remains unclear in the literature whether the Langer lines have a structural basis, and to the authors' knowledge there is no quantitative data published on this point. By identifying the orientation of collagen fibres in the dermis we have shown that there is a correlation between their orientation and that of the Langer Lines. However, it is only when a quantitative study is performed that this becomes apparent. 

Knowledge of the collagen fibre orientation along with the presented stress-strain data will also facilitate the determination of material parameters for structurally based constitutive models for skin. While the test method described here is not sufficient to fully describe the full 3-D response of the material, \citet{Holzapfel06b} has shown how a combination of tensile tests and histological data can be used to reasonably determine the material parameters of a 3D hyperelastic model. In a companion paper \citep{NiAnnaidh11b}, we use a similar approach in order to determine the material and structural parameters of a suitable anisotropic model, the so-called Gasser-Ogden-Holzapfel model \citep{Gasser06a}, readily implemented into the commercial Finite Element code ABAQUS.

\section*{Acknowledgments}
This research was supported by a Marie Curie Intra European Fellowship within the 7$^{th}$ European Community Framework Programme, awarded to MD. It was also supported by IRCSET (Irish Research Council for Science, Engineering and Technology) through the Ulysses Award, and IRCSET and the Office of the State Pathologist (Irish Department of Justice and Law Reform) through the EMBARK scholarship awarded to AN\'{i}A. This work is supported by grants from Ile-de-France.

\nocite{Silver01}
\nocite{Stark77}
\bibliography{destrade-72-final}

\end{document}